\documentclass[pra,twocolumn,amssymb]{revtex4}

\usepackage{amsmath}
\usepackage{amssymb}
\usepackage{bbm}
\usepackage{mathtools}
\usepackage[normalem]{ulem}
\usepackage{dsfont}
\usepackage{mathrsfs}
\usepackage{cancel}

\usepackage{braket}

\usepackage[utf8]{inputenc}
\usepackage[T1]{fontenc}

\def\eq{\text{eq}}

\def\la{\langle}
\def\ra{\rangle}

\def\ii{\mathrm{i}}

\def\bg{\begin{equation}\begin{gathered}}
\def\eg{\end{gathered}\end{equation}}

\def\B#1{\!\left(#1\right)}

\def\be{\begin{equation}}
\def\ee{\end{equation}}

\def\bee{\begin{equation*}}
\def\eee{\end{equation*}}

\def\ba{\begin{equation}\begin{aligned}}
\def\ea{\end{aligned}\end{equation}}

\def\gup{g_{*}^\uparrow}
\def\chiup{\chi_{*}^\uparrow}
\def\gdown{g_{*}^\downarrow}
\def\chidown{\chi_{*}^\downarrow}
\def\eq{{eq}}
\def\reg{{reg}}
\def\sing{{sing}}

\makeatletter
\newcommand{\pushright}[1]{\ifmeasuring@#1\else\omit\hfill$\displaystyle#1$\fi\ignorespaces}
\newcommand{\pushleft}[1]{\ifmeasuring@#1\else\omit$\displaystyle#1$\hfill\fi\ignorespaces}
\makeatother

\usepackage[ddmmyyyy]{datetime}
\makeatletter
\def\Dated@name{}
\makeatother

\begin{document}

\title{
Locating quantum critical points with  Kibble-Zurek quenches
}
\author{Micha{\l}  Bia\l ończyk and Bogdan Damski}
\affiliation{Jagiellonian University, Institute of Theoretical Physics, {\L}ojasiewicza 11, 30-348 Krak\'ow, Poland}
\begin{abstract}
We describe a scheme for finding quantum critical points based on  studies of 
a non-equilibrium susceptibility during finite-rate   quenches 
taking the system from one phase to another. 
We assume that two such quenches are performed in opposite directions, and argue that 
they lead to formation of peaks of a non-equilibrium susceptibility on opposite 
sides of a critical point. 
Its position is then narrowed to the interval marked off by these
values of the parameter driving the transition, at which the
peaks are observed. Universal scaling with the quench time of precision of 
such an estimation is derived and verified in two exactly solvable models. 
Experimental relevance of these results is expected. 
\end{abstract}
\date{\today}
\maketitle

\section{Introduction}
Non-equilibrium phase transitions are ubiquitous in Nature.
Their studies, in the context relevant for this
work, were  started by  Kibble, who investigated 
cosmological phase transitions of the early Universe \cite{KibbleRev}.
It was then  proposed  by Zurek that similar phenomena can be approached in 
tabletop  condensed matter  systems \cite{ZurekRev}.
These theoretical  investigations  triggered experimental 
work on   non-equilibrium dynamics 
of   superconductors, Josephson junctions, superfluids, 
 cold atoms and ions, liquid crystals, multiferroics, convective
fluids, colloids, etc.
Recent  surveys of these efforts, discussing  dynamics of  both classical and quantum
phase transitions,  can be found in  \cite{KibbleToday,JacekAdv2010,PolkovnikovRMP2011,delCampoReva,delCampoRevb}.

Non-equilibrium dynamics, we are interested in, 
comes from finite-rate  driving of a system across its   critical
point. Key features of this process are captured  by the Kibble-Zurek (KZ) theory, which 
relates  non-equilibrium response of a  system  to the quench
rate and some  universal critical exponents.

Phase transitions, however, are also characterized by non-universal   properties,
among which the position of the critical point clearly stands out. 
Indeed, by the very definition, 
it gives us the physical parameter(s) at which the properties of the
system fundamentally  change.
Such a  dramatic  change is possible due to the fact 
that some of the most interesting   many-body
physics takes place near  critical points, where distant parts of the system 
get correlated and its  response to
external perturbations significantly slows down. 
Our understanding of  such phenomena builds on the insights coming from the 
renormalization-group theory, whose basic assumptions are best justified 
very close to critical points \cite{Cardy1}. 
Detailed  studies of these and related phenomena 
cannot proceed  without accurate determination of critical points.
Their knowledge  is also of practical importance, 
which is perhaps best seen in all devices involving superconductors.

\begin{figure}[t]
\includegraphics[width=\columnwidth,clip=true]{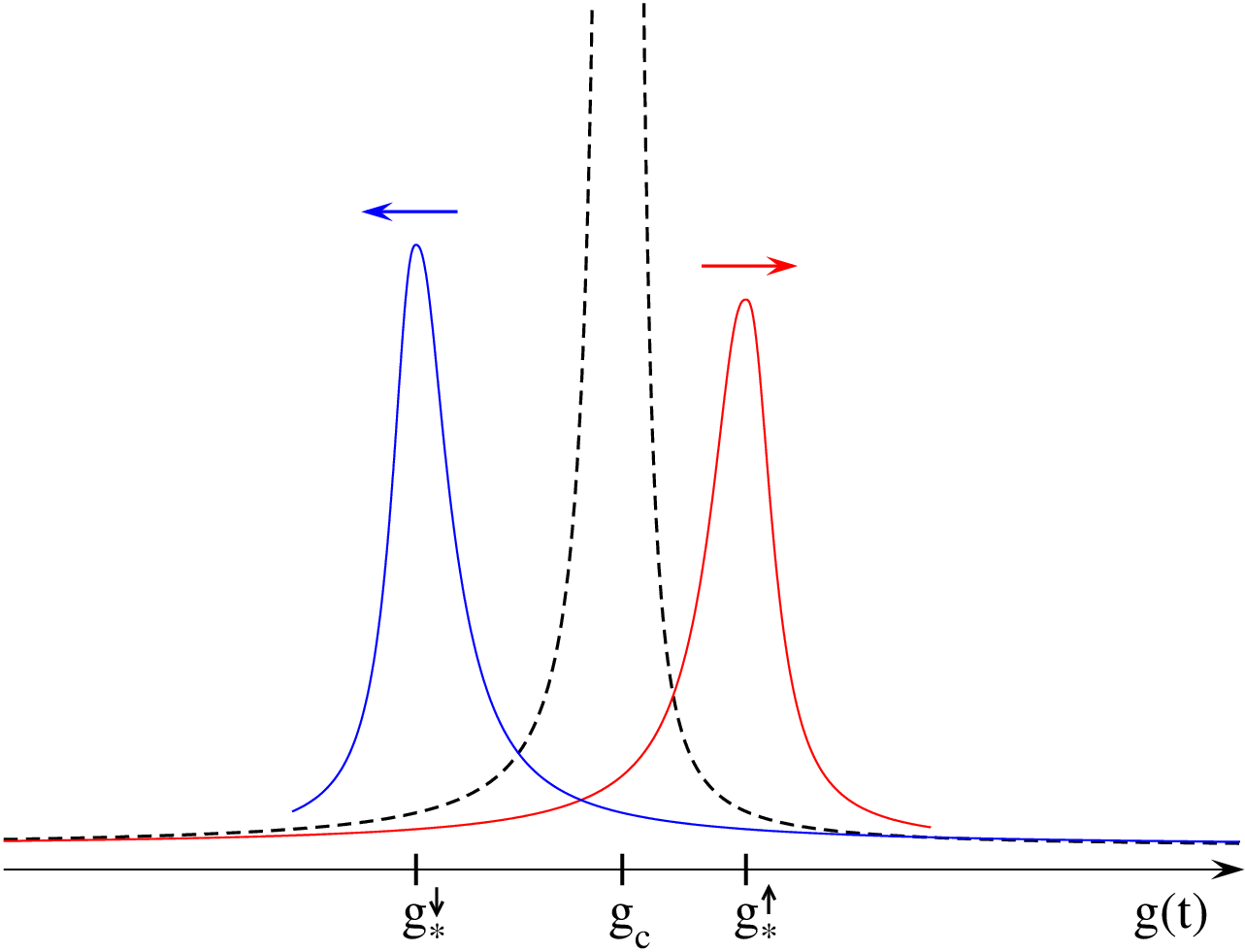}
\caption{Schematic plot of our scheme for determination of the quantum
critical point $g_c$.
Dashed lines show the equilibrium susceptibility, which is divergent at  $g_c$. 
The red (blue) line shows the non-equilibrium susceptibility 
for the quench, where the parameter $g$  is ramped up (down). Arrows
indicate the direction of changes of $g(t)$. The quantum critical point is
supposed to be located between the extrema of these curves, 
$\gdown<g_c<\gup$.
}
\label{schem_fig}
\end{figure}

It is the purpose of this work  to discuss a propitious  KZ-related  scheme for
localization  of quantum critical points (QCPs)--see Fig. \ref{schem_fig} for
its schematic presentation.
So, we will be dealing with 
quantum phase transitions \cite{PiersNature2005,Sachdev,ContinentinoBook,SachdevToday}, 
whose dynamical studies,  in the framework of the quantum KZ theory, were
initiated by \cite{BDPRL2005,DornerPRL2005,JacekPRL2005}. 
Preceding work on quench-based localization of
QCPs can be found in  \cite{ZhongArxiv2013,FanPRB2015,HuangPRB2019}, where 
finite-rate quenches were used, and in \cite{ArnabSciRep2015,ArnabPRB2017}, where  
instantaneous ones were employed. These studies differ from our work in the strategy employed 
for extraction of  QCPs and quench protocols that are used for such
a purpose.

The outline of this paper is the following. We explain the idea behind our
work in Sec. \ref{IdeaSec}. Calculations supporting  it are presented 
in Secs. \ref{ExtendedSec} and \ref{IsingSec}, where respectively extended $XY$ and 
Ising models  are considered. Conclusions and outlook can be found  in Sec. 
\ref{ConclusionsSec}. Technical details, pertinent to studies from 
Secs.  \ref{ExtendedSec} and \ref{IsingSec}, are laid out  in Appendices 
\ref{ExtendedApp} and \ref{IsingApp}, respectively.
Finally, derivation of the scaling ansatz from Sec. \ref{IsingSec} is
discussed 
in Appendix \ref{ScalingApp}.

\section{Idea}
\label{IdeaSec}
To explain the logic behind our work, we consider some susceptibility  $\chi$, whose
ground-state  value $\chi^\eq$ is algebraically divergent at the QCP $g_c$, 
say
\be
\begin{aligned}
&\chi^\eq(g)=\chi^\eq_\reg(g) + \chi^\eq_\sing(g),  \\
&\chi^\eq_\sing(g)\sim|g-g_c|^{-\gamma}, \ \gamma>0,
\end{aligned}
\label{gamma}
\ee
where the regular (singular) at $g_c$ part of $\chi^\eq$ is denoted as
$\chi^\eq_\reg$ ($\chi^\eq_\sing$).

The system will be initially  prepared in a  ground  state far away from the
QCP. It will be then  quenched towards it by linear in time ramp {\it up} 
of the external  parameter  driving the transition 
\be
g(t)=g_c+\frac{t}{\tau_Q},
\label{godt}
\ee
where inverse of the quench time $\tau_Q$ provides the quench rate and the QCP is
reached at the time $t_c=0$.

As long as the system will be far away from the QCP, its evolution will be
adiabatic and so its susceptibility will closely match its instantaneous equilibrium value.
Near the QCP, however,  the evolution  cannot  be adiabatic because
the reaction time of the system, given by the inverse of its
energy  gap \cite{BDPRL2005}, diverges at $g_c$.
So, the  susceptibility 
$\chi(g(t))$  
should  lag behind  $\chi^\eq(g(t))$. 
The mismatch between the two will be largest
at the QCP, where $\chi(g_c)$, unlike $\chi^\eq(g_c)$, will
be finite (no singularities are expected in the non-equilibrium state of
the system as it will not be given enough time to develop
them). Another consequence of the delayed reaction to crossing of the QCP should be 
seen in the maximum of $\chi(g(t))$, which we expect to  appear past the QCP, 
say at $\gup>g_c$. 

Suppose now that the system is initially prepared in a ground  state on
the other side of the transition, and    the parameter  
$g(t)$ is ramped {\it down}. The same  discussion then leads to the conclusion
that $\chi(g(t))$  
 should have the maximum at some
$\gdown<g_c$. Therefore, we expect that  location of
the QCP can be pinned down to the interval $(\gdown,\gup)$. 

This qualitative description can be made quantitative with the
KZ theory, which introduces 
the characteristic non-equilibrium   time scale 
$
\hat t\sim \tau_Q^{z\nu/(1+z\nu)}
$
and the interrelated field  scale
\begin{align}
\label{ghat}
&\hat g=|g(t_c\pm\hat t\,)-g_c|\sim\tau_Q^{-1/(1+z\nu)}, 
\end{align}
where $z$ and $\nu$ are the  dynamical and correlation-length universal critical exponents.
It can be then argued that near  the QCP, the non-equilibrium
susceptibility $\chi$ will be dominated, for slow-enough quenches,
by its
universal part 
\be
\begin{aligned}
&\chi(g(t)) \approx \hat\chi f\B{\frac{g(t)-g_c}{\hat g}},\\
&\hat\chi=\chi^\eq[g(t_c\pm\hat t\,)]\sim\tau_Q^{\gamma/(1+z\nu)},
\end{aligned}
\label{chit}
\ee
where  $f(x)$ is a non-singular  scaling function, which  is proportional to $|x|^{-\gamma}$
 before the onset of 
non-equilibrium dynamics, so that $\chi(g(t))\approx\chi^\eq_\sing(g(t))$ 
there.  
Ansatz (\ref{chit}) combines two basic ingredients of the KZ theory. First, the
adiabatic-impulse approximation 
assuming 
that system's dynamics is frozen in the impulse regime, i.e. when 
$|t-t_c|<\hat t$, and adiabatic before 
entering it \cite{BDPRL2005,BDPRA2006,ZanardiPRA2018}. 
This introduces $\hat\chi$ into (\ref{chit}). Second,
the assumption that non-equilibrium dynamics of physical observables should depend
on the rescaled time difference $(t-t_c)/\hat t$, which explains the scaling function in (\ref{chit}). The latest
take on this ansatz  can be found in 
\cite{KolodrubetzPRL2012,SondhiPRB2012,FrancuzPRB2016,DebasisPRB2020,VicariPRR2020}, see e.g. 
\cite{BDPRL2010} for preceding work in the context  of
classical  phase transitions.

It now follows from  (\ref{chit}) that precision of  QCP
determination should   increase  with the quench time as
\be
\delta=\gup-\gdown\sim   \tau_Q^{-1/(1+z\nu)}.
\label{delta}
\ee
Several remarks are in order now. 

First, we propose that the above-outlined scheme  can be used for either numerical or
experimental localization of QCPs. 

Second,  arguments presented between (\ref{gamma}) and (\ref{ghat}) are based on general 
considerations, which do not involve the KZ theory. For this reason, they should be presumably
more robust than KZ predictions, which are laid out in   (\ref{chit}) and
(\ref{delta}). So, even in systems where quantitative 
verification of the KZ theory is challenging,  our
technique for localization of QCPs may still be  useful. 

Third, it is interesting to realize that 
just a single, in each direction, 
sweep  of the parameter driving the transition may be sufficient for
reasonably-accurate estimation of the position of the QCP.
Note that  one cannot get 
both upper and lower bounds on the 
position of the QCP from a single one-way quench.
Our two-way quench protocol gets around this limitation.

Fourth, our scheme does not specify,
where the QCP is located between the maxima of non-equilibrium 
susceptibilities, i.e. within the interval $(\gdown,\gup)$.
The fact, that the position of the QCP is supposed to be bounded 
in such a way, could be  of  practical relevance,
because  features such as
extrema are typically the easiest to  extract from experimental data.
However, if determination  of some susceptibility would require differentiation of
such presumably noisy data, one will have  to smooth that data first. 
This  can be done with various
   techniques, see e.g. \cite{BDSciRep2019} for the Pad\'e approximant example.


Fifth, result (\ref{delta})  is of interest from the metrological perspective
and it is worth to stress that 
the KZ theory has been  only recently systematically
explored  in the metrological context
\cite{MarekPRX2018}. Moreover, (\ref{delta})  can be also used for extracting  
the product of universal critical exponents.

Two exactly solvable models will be used below  for illustration of 
above-introduced concepts. Their numerical solutions will be presented on 
Figs. \ref{chitXY}--\ref{rescalIsing}. Technical details of our simulations
can be found in Appendices \ref{IsingApp} and \ref{ExtendedApp}.

\begin{figure}[t]
\includegraphics[width=\columnwidth,clip=true]{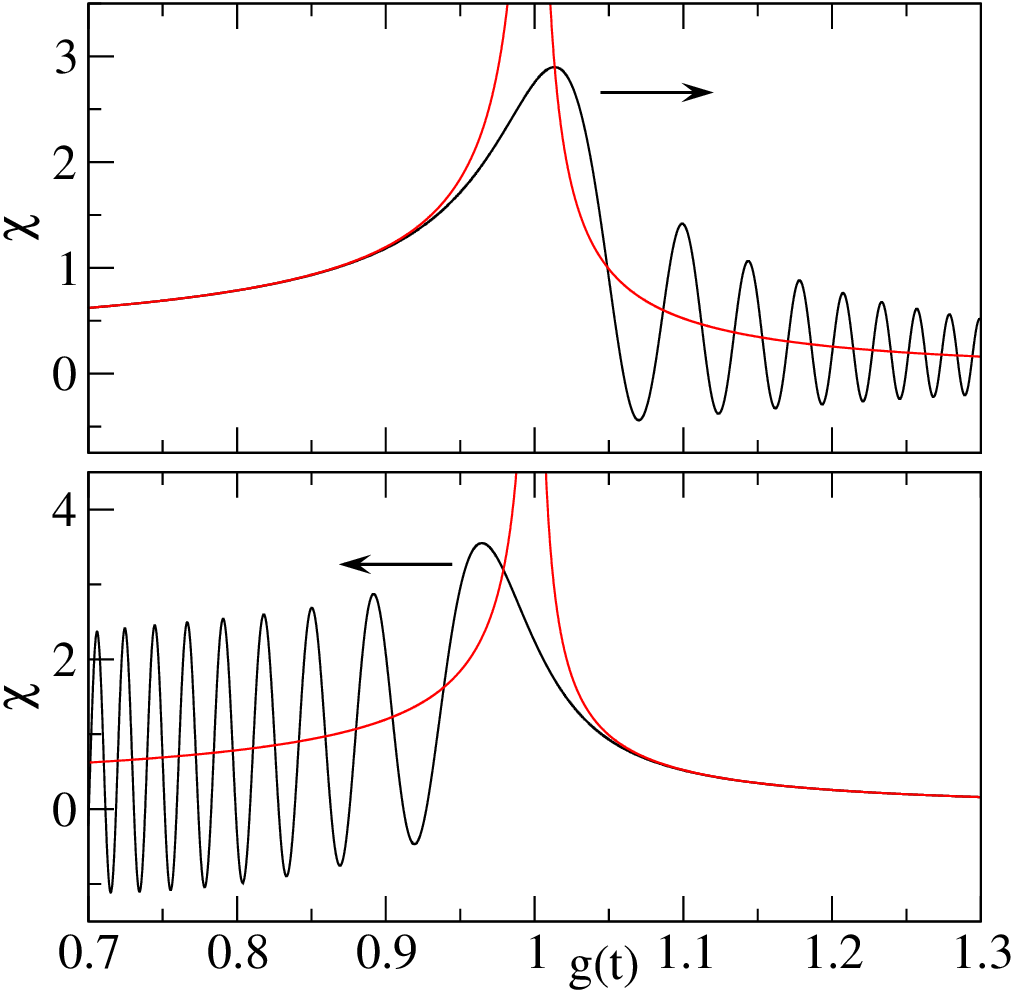}
\caption{
Dynamics of the susceptibility $\chi(g(t))$ during up and down 
quenches in the extended $XY$ model is presented by black lines
in  upper and lower panels, respectively. The instantaneous 
equilibrium value of the susceptibility, $\chi^\eq(g(t))$, 
is provided by  red lines. The quench time $\tau_Q=300$.
}
\label{chitXY}
\end{figure}

\section{Extended $XY$ model}
\label{ExtendedSec}
We take   the Hamiltonian
\begin{equation}
H = -\sum_{i=1}^N (\sigma_i^x \sigma_{i+1}^x +  \tfrac{1}{3}\sigma_i^y \sigma_{i+1}^y 
-\tfrac{1}{3} \sigma_i^x \sigma_{i+1}^z \sigma_{i+2}^x  + g \sigma_i^z),
\label{HEXY}
\end{equation}
where $g\ge0$ is the external magnetic field, 
$\sigma^{x,y,z}_i$ are Pauli matrices acting on the $i$-th spin, 
$N$ is the number of spins, and periodic  boundary conditions are implemented. 
Basic 
properties of this model  were described in \cite{SuzukiProgTeorPhys1971,DebasisPRB2020}.
Its  QCP is at  $g_c=1$ and it separates  ferromagnetic  ($0<g<1$) and
 paramagnetic  ($g>1$) phases. Critical exponents of (\ref{HEXY}) 
are $z=3$ and $\nu=1/3$.

As the susceptibility of interest in our translationally-invariant system, we take
the derivative of the transverse
magnetization at an arbitrary lattice site 
\be
\chi=\frac{d\la\sigma^z_i\ra}{dg}.
\label{chi}
\ee

In equilibrium, this quantity is algebraically divergent at the
QCP. 
Indeed, $\chi^\eq=-d^2E_0/dg^2$  from 
the Feynman-Hellmann theorem,
where $E_0$ is the ground state energy per lattice site. The singular part of $E_0$ is typically
assumed to scale as $|g-g_c|^{2-\alpha}$. If we now combine this insight with 
the quantum hyperscaling relation--i.e.  $\alpha=2-\nu(d+z)$,
where $d$ represents system dimensionality \cite{ContinentinoBook}--we will get that 
$\alpha=2/3$ for model (\ref{HEXY}). Thus,   $\chi^\eq_\sing(g)$ is given by (\ref{gamma}) 
with $g_c=1$ and $\gamma=2/3$, which 
we have numerically verified, and so $\hat\chi\sim
\tau_Q^{1/3}$.

\begin{figure}[t]
\includegraphics[width=\columnwidth,clip=true]{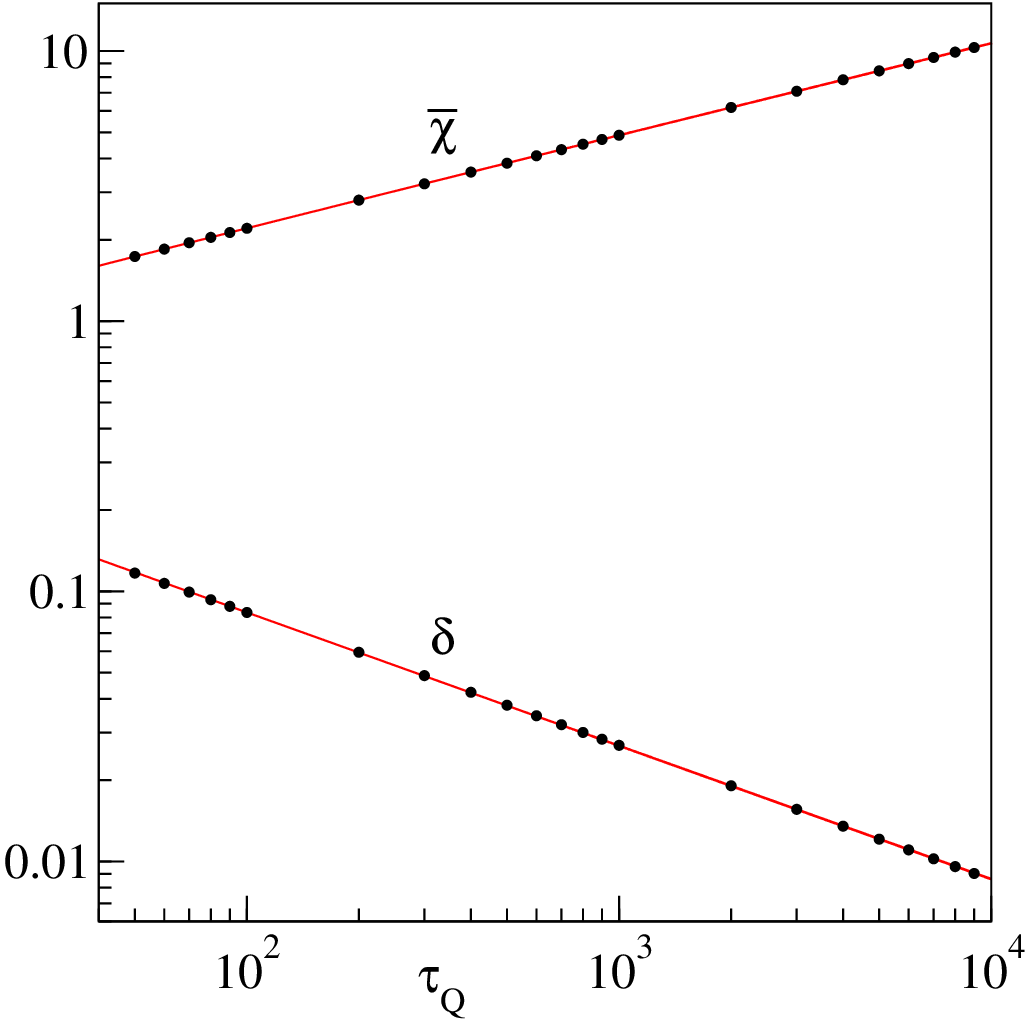}
\caption{
Global maxima of the non-equilibrium  susceptibility $\chi$ of the 
extended $XY$  model. 
Upper data: 
the average value of $\chi$ at global
maxima  during up and down quenches.   
Lower data: the  distance $\delta$ between such maxima. 
Red lines show fits to black dots coming from numerics, see (\ref{lnXY}) and (\ref{chidXY}).
}
\label{deltaXY}
\end{figure}

The ramp up of the magnetic field will be done with 
\be
g(t)=\B{\frac{t}{2\tau_Q}}^2,
\label{gupt}
\ee
while its ramp down will be done with
\be
g(t)=g_0-\frac{t}{\tau_Q}, \ g_0=5.
\label{gdownt}
\ee
Both quenches start from ground states at $t=0$ and then the system is driven towards the
QCP, which is  reached at $t_c$ equal to  $2\tau_Q$ and 
$(g_0-1)\tau_Q$ for  up  (\ref{gupt})  and down (\ref{gdownt})   quenches, respectively.

These quenches share the same important property. Namely, their 
rate, given by $|dg/dt|$, is  equal to $\tau_Q^{-1}$ at $t_c$. 
Thus, near the QCP,  the driving  proceeds 
just as  in  model quench (\ref{godt}), which 
is all that should matter in the context of the KZ theory
(see e.g. \cite{BDNJP2008} for a similar quadratic-in-time quench studied
in the KZ framework). 

\begin{figure}[t]
\includegraphics[width=\columnwidth,clip=true]{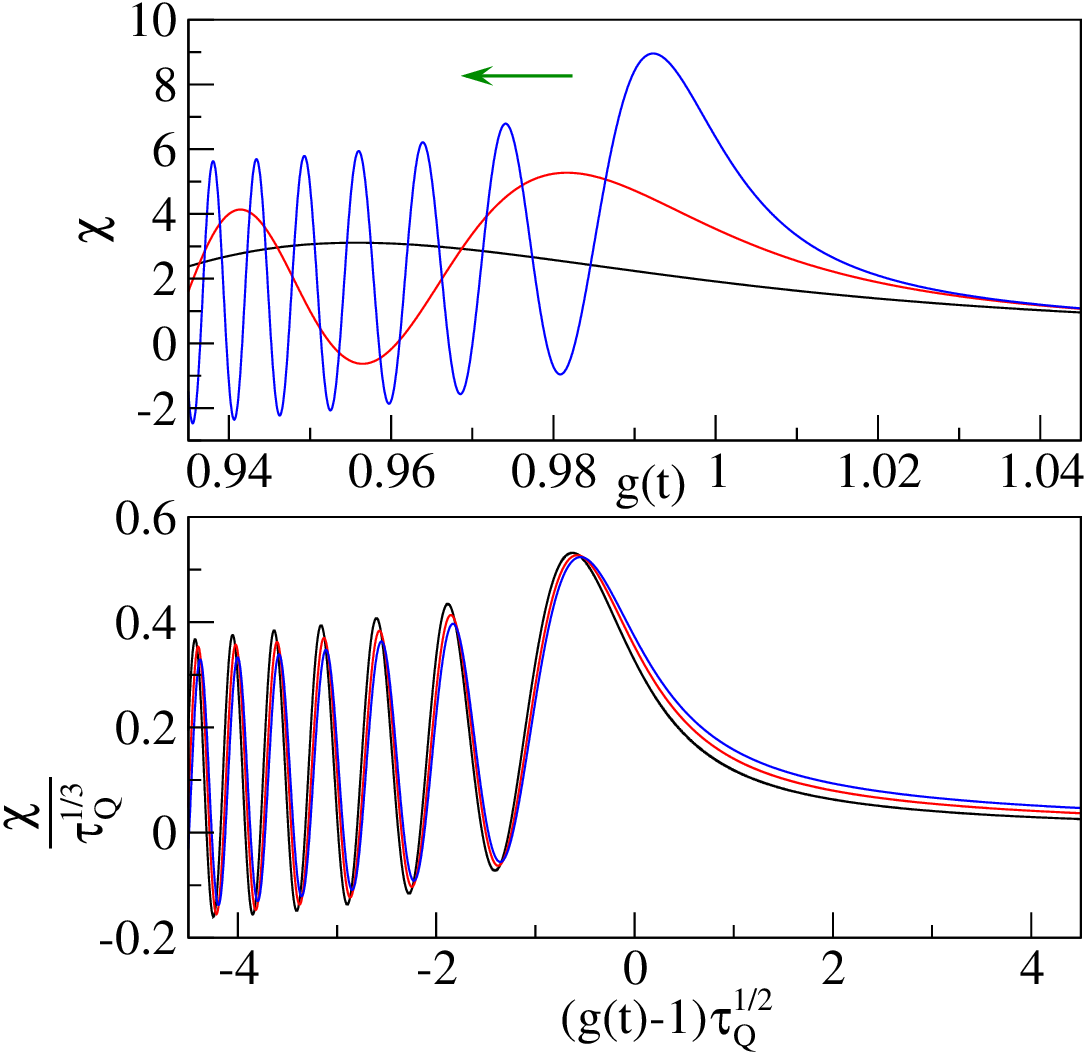}
\caption{
Dynamics of the susceptibility $\chi$ during down quenches of the extended $XY$ model. 
Black, red, and blue lines are obtained  for $\tau_Q=200$, $1000$, and $5000$,
respectively. Panels show results before and after  KZ rescalings. 
}
\label{rescalXY}
\end{figure}

As for  differences between (\ref{gupt}) and (\ref{gdownt}), 
we note that quenches typically produce excitations at the beginning of time evolution 
(see e.g. \cite{MarekArxiv}).
Such excitations are of no interest in our studies.
They appear because  variations  of  the external parameter are not  smoothly turned on
(the more low-order
 derivatives of the external parameter vanish  at $t=0$, the more adiabatic the quench initially is).
Quadratic time dependence in (\ref{gupt}) 
noticeably reduces  initial  excitation of our system with respect to what 
would happen if the up quench would be linear. 
The situation is somewhat similar for down quenches (\ref{gdownt}). 
However, the gap in
the excitation spectrum 
is much larger at $g_0\gg1$  than at $g=0$. Even for linear 
quenches, we find that it sufficiently reduces initial  non-adiabaticity of
our observable during  ``down'' evolutions (the same happens in the Ising model studied
in Sec. \ref{IsingSec}).

Typical dynamics of  susceptibility (\ref{chi})  is presented in 
Fig. \ref{chitXY}, where we see three distinct regimes.
First, the evolution is adiabatic. Then, near the QCP,  the susceptibility 
lags behind its instantaneous equilibrium value,  and a well visible global maximum appears after crossing the 
QCP.
Finally,  the susceptibility oscillates, which 
can be regarded  as a quasi-adiabatic stage (no more excitations
are generated,  system's dynamics revolves around the instantaneous
equilibrium solution).

More quantitatively, from numerical data  presented in  Fig. \ref{deltaXY}, we find that 
the distance $\delta$ between  global maxima for up and down quenches is described by
\be
\ln\delta=-0.209(3)-0.4936(5)\ln\tau_Q.
\label{lnXY}
\ee
This result comes  from a linear regression \cite{Regression}. 
It is  in excellent
agreement with (\ref{delta}) suggesting a prefactor of $-1/2$ in front of the
logarithm.

It is also instructive to have a look at the average value of the
susceptibility at  global maxima, 
which we denote by $\chiup$ and $\chidown$ for up and down quenches,
respectively. 
The nonlinear fit to data from Fig. \ref{deltaXY} shows that  
\be
\overline{\chi}=(\chiup+\chidown)/2=-0.063(2) + 0.4795(7)\tau_Q^{0.3376(2)}, 
\label{chidXY}
\ee
where the exponent is in excellent agreement with the value of $1/3$
due to $\gamma=2/3$ and $z\nu=1$ (\ref{chit}). 
Similar results are obtained
when the fit is individually performed for 
either $\chiup$ or $\chidown$. Computation of $\overline{\chi}$, however, removes 
small deviations  of $\chiup$ and $\chidown$ from the perfect   KZ scaling
solution. Those deviations  are presumably caused by  non-universal contributions to
susceptibilities 
(see lower panel of Fig. \ref{rescalXY}, where collapse of  global 
 maxima, after KZ rescalings, is not exact and note that the KZ theory overlooks
  non-universal dynamics).
Analogical remarks apply to the discussion of $\overline{\chi}$ in Sec.
\ref{IsingSec}, and so they will not be repeated there.

Finally, we mention that   ansatz (\ref{chit}) is verified in Fig. \ref{rescalXY}, where
good overlap between non-equilibrium susceptibilities, obtained for vastly
different quench times $\tau_Q$, is seen  after proper rescalings.

\begin{figure}[t]
\includegraphics[width=\columnwidth,clip=true]{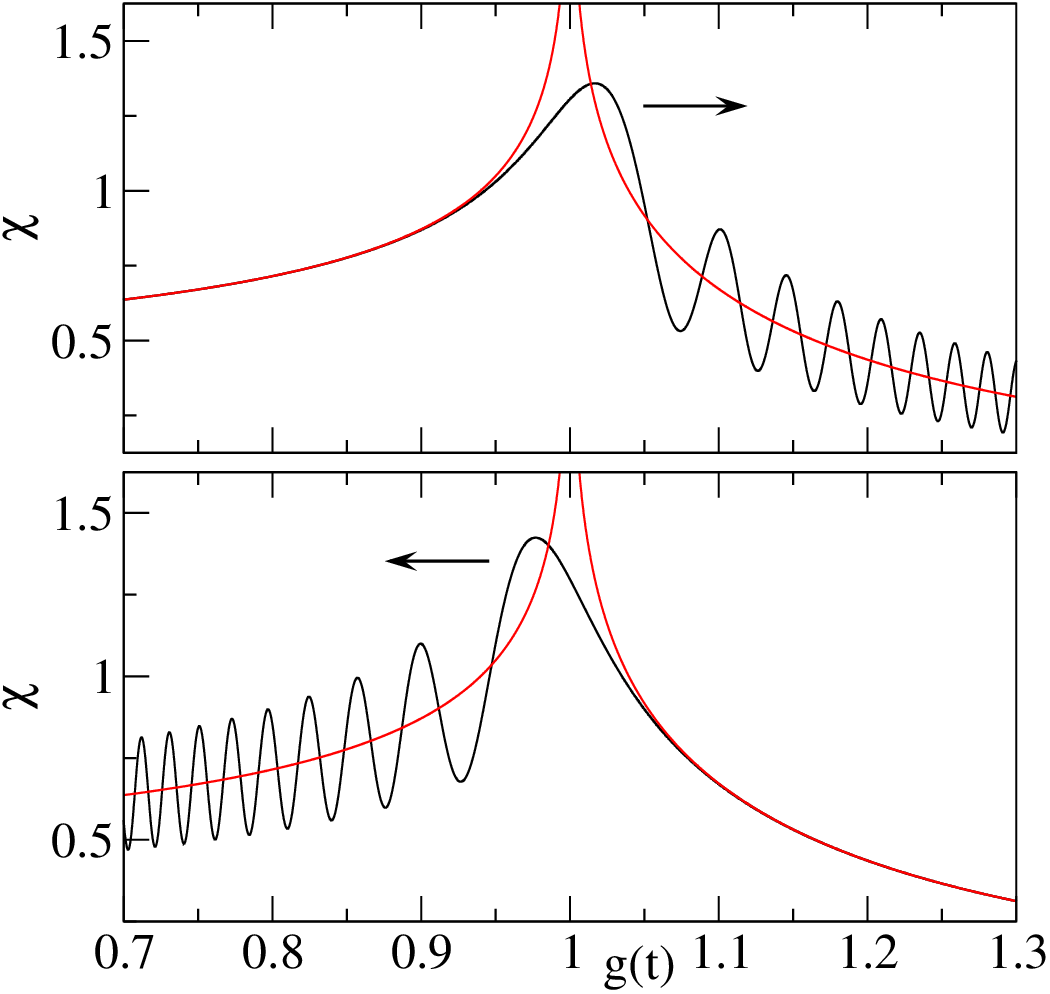}
\caption{Black lines: the susceptibility $\chi(g(t))$ during up (upper panel) and
down (lower panel) quenches of the Ising model with  $\tau_Q=300$. Red lines: 
the instantaneous equilibrium value of $\chi$.
}
\label{chitIsing}
\end{figure}

\section{Ising model}
\label{IsingSec}
The Hamiltonian of interest
now is 
\be
H = -\sum_{i=1}^{N}(\sigma^x_i\sigma^x_{i+1} + g \sigma^z_i),
\label{HIsing}
\ee
where the QCP and phases are the same as in the 
extended $XY$ chain \cite{Lieb1961,Pfeuty,BDJPA2014}.
  Dynamics of this paradigmatic 
 model, under continuous driving such as (\ref{godt}), 
 was studied in \cite{JacekPRL2005,DornerPRL2005,PolkovnikovPRB2005,RalfPRA2007,PolkovnikovPRL2008,SenPRB2009,SantoroPRB2009,JacekPRA2007,SenPRA2009,ArnabPRB2010,KolodrubetzPRL2012,ZhongArxiv2013,Dutta2015,FrancuzPRB2016,SantoroJstat2015,BDproceedings,PuskarovSciP2016,ApollaroSciRep2017,MichalOne,AdolfoPRL2018,MarekArxiv,MichalTwo}. 
 Key differences between (\ref{HEXY}) and (\ref{HIsing})  are seen through
critical exponents, which are now given by  $z=\nu=1$. 

\begin{figure}[t]
\includegraphics[width=\columnwidth,clip=true]{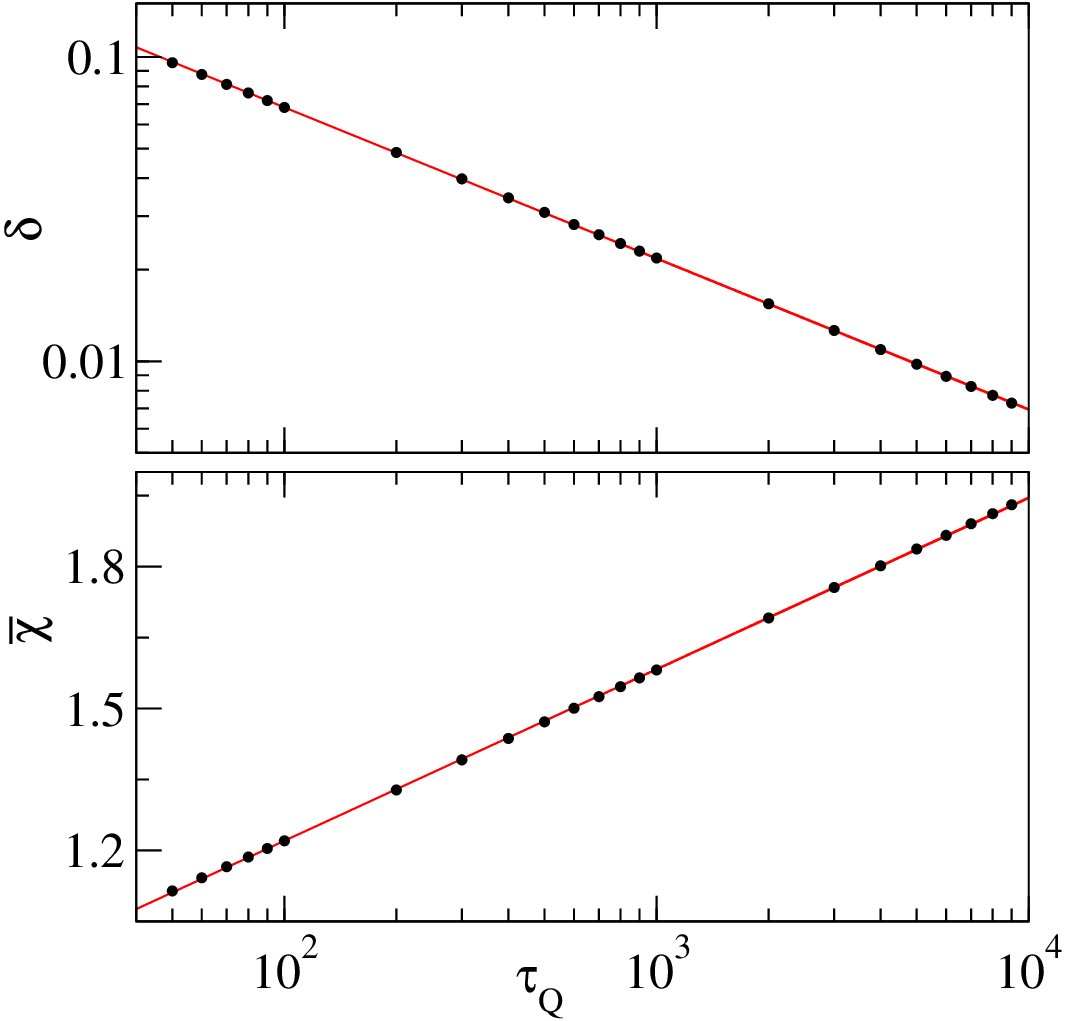}
\caption{Upper panel: the distance $\delta$ 
between global maxima of the susceptibility $\chi$ appearing during up and down quenches of the Ising model.
Lower panel: arithmetic average of the value of $\chi$ at those maxima.
Black dots show numerics, red lines are fits (\ref{lnIsing}) and
(\ref{chiuIsing}). 
}
\label{deltaIsing}
\end{figure}

Such values lead to 
non-algebraic  singularity of the equilibrium version of susceptibility (\ref{chi}), which can be 
understood by noting that $\alpha=0$ now. 
Indeed, $\chi^\eq_\sing$  has the following expansion  near the QCP \cite{LackiJSTAT2017}
\be
\chi^\eq_\sing(g)\approx -\frac{1}{\pi}\ln|g-1|.
\label{eqIsing}
\ee
This well-known result is a bit
unusual because  one typically expects algebraic singularities as in  (\ref{gamma}).
Logarithmic singularity of  $\chi^\eq_\sing$
has  interesting consequences on $\chi$, whose scaling properties are not  captured by
(\ref{chit}). The appropriate KZ ansatz reads 
\be
\label{lnchit}
\chi(g(t)) \approx f\B{\frac{g(t)-g_c}{\hat g}} + \frac{\ln\tau_Q}{2\pi},
\ee
where  up to a constant term, $f(x)$ is approximated by $-\frac{1}{\pi}\ln|x|$
before the onset of non-equilibrium dynamics. 
To derive   (\ref{lnchit}),  we start with $d^2\la\sigma^z_i\ra/dg^2$, whose equilibrium value is
algebraically  divergent at the QCP, apply  ansatz
(\ref{chit}) to it,  integrate the resulting expression, and take into
account adiabaticity before the beginning of non-equilibrium dynamics
(see Appendix \ref{ScalingApp} for a detailed discussion). 
Alternatively, one may adopt results from \cite{LackiJSTAT2017}, where $d\la\sigma^z_i\ra/dg$ was studied
with renormalization group techniques in a time-independent  but spatially inhomogeneous Ising chain. 
This is done by replacing $\lambda_Q$ with $\tau_Q$ in Eq. (26)  from \cite{LackiJSTAT2017}.
Finally, it may be also worth to mention that, to the best of our knowledge,
 scaling ansatz (\ref{lnchit}) has never been applied to time quenches before.

\begin{figure}[t]
\includegraphics[width=\columnwidth,clip=true]{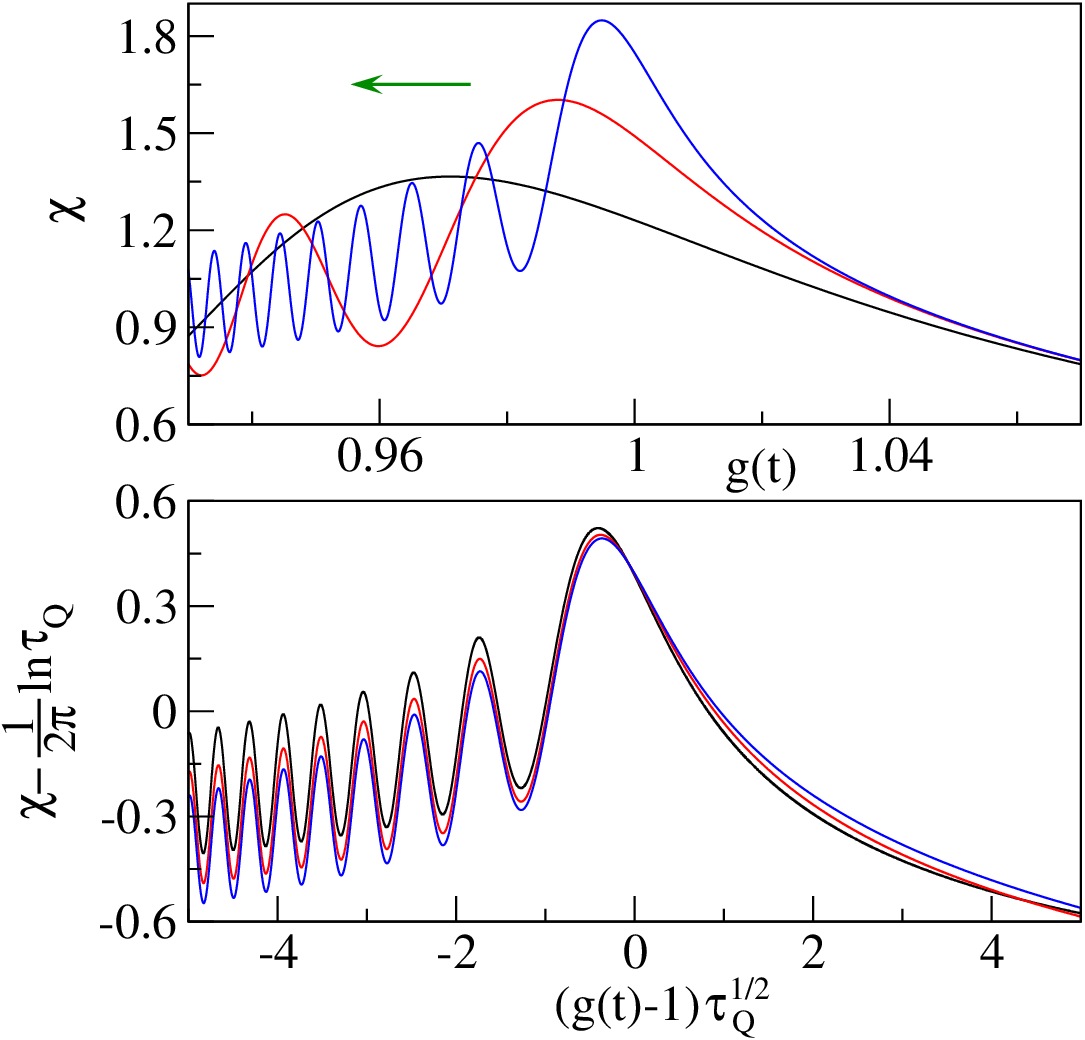}
\caption{
Dynamics of the susceptibility $\chi$ during  down quenches of the Ising
model: before and after  KZ rescalings motivated by ansatz (\ref{lnchit}).
Black, red, and blue lines are obtained  for $\tau_Q=200$, $1000$, and $5000$.
}
\label{rescalIsing}
\end{figure}

Typical dynamics of the susceptibility $\chi$, due to   either (\ref{gupt}) or (\ref{gdownt}),
is shown in Fig. \ref{chitIsing}. The fits from Fig. \ref{deltaIsing}  are
\begin{align}
\label{lnIsing}
&\ln\delta=-0.397(3)-0.4965(4)\ln\tau_Q,\\
\label{chiuIsing}
&\overline{\chi}=0.495(2)+0.1575(2)\ln\tau_Q, 
\end{align}
which can be compared to our  theory. The prefactor in front
of the logarithm in (\ref{lnIsing}) should be $-1/2$, and indeed it is very
much so. The one in  (\ref{chiuIsing}) is  also very close to our expectations, 
i.e. $1/2\pi\approx0.159$ due to (\ref{lnchit}). 
Finally, verification  of ansatz  (\ref{lnchit}) is shown in Fig.
\ref{rescalIsing}, where pretty good overlap between curves is found.

\section{Conclusions}
\label{ConclusionsSec}
Summarizing, we have proposed how quantum critical points can be 
accurately localized by scanning a non-equilibrium susceptibility during  Kibble-Zurek
quenches. Our scheme assumes that   two such scans are 
performed 
by either increasing or decreasing the external parameter driving the
transition. We have argued that each of them should  lead to formation  of 
a  peak  of a  non-equilibrium susceptibility, and that 
the critical point can be pinned down to the interval 
marked off by  these  values of the external parameter, at which 
the peaks  are observed. The width of such an interval  has been argued to exhibit 
universal power-law scaling with the quench time, shrinking to zero in the
adiabatic limit. 

We have tested these predictions in two exactly solvable models, exhibiting 
either algebraic or logarithmic singularities 
of  equilibrium susceptibilities.
We have found that each quench actually produces a train 
of progressively smaller susceptibility peaks. By focusing on the
highest one for each quench, our  expectations have been 
precisely confirmed.  
There are several prospective extensions of these 
studies. 

First,  similar calculations can be done  in 
 other  models. 
 This should   increase  understanding of   susceptibilities during
 Kibble-Zurek quenches. 
This is arguably a  poorly explored topic as  we have been able to find only three references
 exploring susceptibilities in the Kibble-Zurek context 
 \cite{ZhongArxiv2013,LackiJSTAT2017,MarekArxiv}. Out of them,   
\cite{LackiJSTAT2017} is not even  focused on non-equilibrium dynamics of the kind we discuss in this work
as it deals with spatial Kibble-Zurek quenches \cite{TurbanJPA2007,DornerPRS2008}. 
It should be also noted that as long as $\tau_Q<\infty$, there will be
always non-universal  contributions to dynamics of 
susceptibilities, which  are not captured by the Kibble-Zurek theory.
Their quantification requires system-specific studies.

Second,  our scheme can be
 used for numerical localization of quantum critical points in non-exactly
 solvable models, providing a complementary approach  to studies based on evaluation 
of equilibrium susceptibilities. This  complementarity
can be seen by noting that generation of quantum states,
used for computation  of  equilibrium (non-equilibrium) susceptibilities,  can
be done by imaginary (real) time evolutions. 
It can be also seen  from  the perspective
 of  tensor network simulations 
 \cite{VerstraeteAdvPhys2008,SCHOLLWOCKAnnals2011,OrusAnnals2014}, where equilibrium calculations can be done
 with variational  methods, while the non-equilibrium ones, needed for
 exploration of our scheme,  would follow  from their time-dependent extensions.

Third, we expect that our predictions
could be used for  experimental determination of phase diagrams 
 of physical systems through their non-equilibrium response to variations of
 external fields. They could also  motivate experimental studies of
  susceptibilities during Kibble-Zurek quenches.
 In particular, one should be able to 
 implement and test our scheme in cold atom and ion 
emulators of spin systems \cite{PorrasPRL2004,KorenblitNJP2012,LewensteinAdv,BlochSci2017,SchaussQST2018}, whose
dynamical experimental studies were recently reported in 
\cite{LukinNature2017,MonroeNature2017,LukinNature2019,LMGexp}.
The important  open question here is how robust our approach is to 
environmental couplings    and ``imperfections'' of  real
experimental setups. 

Finally, we would also like to mention
that similar phenomena  may be also noticeable during  
non-equilibrium  classical phase transitions.
We think so  because the Kibble-Zurek theory similarly describes dynamics of  quantum and classical
systems
\cite{KibbleToday,JacekAdv2010,PolkovnikovRMP2011,delCampoReva,delCampoRevb}.
In fact, it was originally developed in the classical context  \cite{KibbleRev,ZurekRev}.
Thus, extension of our studies to various systems undergoing classical
phase transitions, such as
those listed  at the very beginning of this work, looks to us like a promising
research direction.

\section*{ACKNOWLEDGEMENTS}

We thank Adolfo del Campo for a remark triggering  our interest in this
subject and for his comments about the manuscript. We also thank 
Marek Rams for  discussions of the extended $XY$ model and his remarks about
the manuscript.
MB and BD were supported by the Polish National Science Centre (NCN) grant DEC-2016/23/B/ST3/01152.

\appendix

\section{Numerical simulations of Ising model}
\label{IsingApp}
We will outline here basic steps leading to  efficient
numerical simulations of  periodic Ising model (\ref{HIsing}).

To begin, we note that  Hamiltonian (\ref{HIsing})  commutes with the parity
operator $\prod^N_{i=1}\sigma^z_i$, whose eigenvalues are  $\pm1$. 
This leads to splitting of the Hilbert space into positive- and
negative-parity subspaces, 
where  eigenstates of $H$ have either $+1$ or $-1$ parity.
Moreover, the parity of the system's state is preserved during time evolutions.
Our evolutions start from ground states in the positive-parity subspace.
Moreover, we consider systems composed of an even number of spins,
which simplifies a bit the following discussion  (see \cite{BDJPA2014} for a
comprehensive discussion of the impact of the parity and  system size on
the spin-to-fermion mapping that we employ below).

The following discussion is 
conveniently carried out  by mapping spins onto  non-interacting fermions via the Jordan-Wigner
transformation
\be
\begin{aligned}
&\sigma^z_i=1-2 c_i^\dag c_i, \ \sigma^x_i=( c_i+ c_i^\dag)\prod_{j<i}(1-2 c_j^\dag c_j),\\
&\{c_i, c_j^\dag\}=\delta_{ij}, \ \{c_i, c_j\}=0,
\end{aligned}
\label{SJW}
\ee
where  anti-periodic boundary conditions have to be  imposed on the fermionic 
operators $c_i$ because we work in the positive-parity subspace \cite{JacekPRL2005,BDJPA2014}.

One then goes to the  momentum space through the substitution 
\be
\begin{aligned}
&c_j= \frac{\exp(-\ii\pi/4)}{\sqrt{N}}\sum_{K=\pm k} c_K\exp(\ii K j),\\
&k=\frac{\pi}{N},\frac{3\pi}{N},\cdots,\pi-\frac{\pi}{N},
\end{aligned}
\ee
arriving at 
\be
\begin{aligned}
 H = 
2\sum_k \big[& ( c_k^\dag  c_k - c_{-k}  c_{-k}^\dag) (g-\cos k)\\
+&  ( c_k^\dag c_{-k}^\dag +  c_{-k} c_{k})\sin k\big],
\end{aligned}
\label{SHk}
\ee
which can be diagonalized via the Bogolubov transformation. The ground state of (\ref{SHk}) is
\begin{align}
& |g\ra=\prod_{k} 
 (u_k^\eq - v_k^\eq c_k^\dag c_{-k}^\dag)|{\rm vac}\rangle, \\
 \label{uveq}
&u_k^\eq=\cos\frac{\theta_k}{2}, \ v_k^\eq=\sin\frac{\theta_k}{2}, \\
\label{thetaknew}
&\sin \theta_k = \frac{\sin k}{\sqrt{g^2-2g\cos k+1}}, \\ 
&\cos\theta_k= \frac{g-\cos k}{\sqrt{g^2-2g\cos k+1}},
\label{thetak}
\end{align}
where 
the vacuum state $|{\rm vac}\rangle$ is annihilated by all $c_{\pm k}$ operators.

From the  time-dependent Schr\"odinger equation, 
\be
\ii\frac{d}{dt}|\psi(t)\rangle= H[g(t)]|\psi(t)\rangle,
\ee  
one then finds that \cite{JacekPRL2005}
\begin{align}
&|\psi(t)\rangle =\prod_{k} (u_k(t) - v_k(t) c_k^\dag c_{-k}^\dag)|{\rm vac}\rangle,\\
&\ii\frac{d}{dt}
\left(
\begin{array}{c}
v_k \\
u_k
\end{array}
\right)
=
2\left(
\begin{array}{cc}
g(t)-\cos k & -\sin k\\
-\sin k  & \cos k-g(t)
\end{array}
\right)\left(
\begin{array}{c}
v_k \\
u_k
\end{array}
\right).
\label{Sidt}
\end{align}

The above  differential equations are efficiently numerically solved with standard techniques (we use the 
Bulirsch-Stoer method \cite{Recipes}). 
The initial conditions are chosen such that  $u_k$ and $v_k$, at the beginning of time evolution, are equal to 
$u_k^\eq$ and $v_k^\eq$, respectively. The latter are
computed at the initial value of the magnetic field.

The non-equilibrium  transverse magnetization is given by
\be
S^z(g(t))=\la\psi(t)|\sigma^z_i|\psi(t)\ra=1-\frac{4}{N}\sum_k |v_k(t)|^2.
\label{Szxcvb}
\ee
Its equilibrium value is obtained after the  replacements $|\psi(t)\ra\to|g\ra$ and  $v_k(t)\to v^\eq_k$.
From (\ref{Szxcvb}), the non-equilibrium susceptibility $\chi$ is numerically computed via
\be
\chi(g_{i+1/2})=\chi\B{g_i+\frac{g_{i+1}-g_i}{2}}
\approx 
\frac{S^z(g_{i+1})-S^z(g_i)}{g_{i+1}-g_i}, 
\ee
where $g_i=g(t_i)$.
The grid, which we use for computation of this quantity, is   $g_{i+1}-g_i=5\cdot10^{-5}$.
The equilibrium susceptibility $\chi^\eq$ is trivially computed through analytic
differentiation. 

We identify positions of maxima of the non-equilibrium susceptibility in the
following way. First, we choose $\tau_Q$ and find  the global  maximum
on the susceptibility vs. magnetic field plot, say at 
$(g_{j+1/2},\chi(g_{j+1/2}))$. We then fit
a parabola to the points $(g_{i+1/2},\chi(g_{i+1/2}))$ around it, satisfying
$\chi(g_{i+1/2})\ge\chi(g_{j+1/2})(1-\epsilon)$, where $\epsilon=0.0025$ is chosen 
(the twice larger $\epsilon$
gives essentially  identical results). The maximum of
such obtained parabola is then analyzed  in the main body of this paper. 
  The fitting
procedure makes our results independent of tiny oscillations  of data
points. It also allows for interpolation of  positions of  maxima
between the grid points.

Our numerics, presented in the main text, 
 have been 
done for  systems composed of $N=2000$ spins. This
imposes an upper limit on quench times, for which Kibble-Zurek dynamics should be
free from finite-size effects. Namely,  the size of the system should be much larger than 
the  correlation length around the time, when the system goes out of
equilibrium  \cite{JacekAdv2010}.
The latter  is proportional to  
$\tau_Q^{\nu/{(1+z\nu)}}$ \cite{DornerPRL2005,JacekAdv2010}.
So, this condition leads to $\tau_Q\ll N^2$ in the Ising chain, which is satisfied in all our numerical 
simulations. In accordance with these expectations, 
we have directly verified that virtually identical results, 
to those  reported in the main text,
are also obtained when $N=1000$. Finally, in the main body of our work, 
we do the fits to numerics  in the range $50\le\tau_Q\le9000$.

\section{Numerical simulations of  extended $XY$ model}
\label{ExtendedApp}
The  procedure, leading to  efficient
numerical simulations of  periodic extended $XY$  model (\ref{HEXY}),
is
similar to the one discussed in Appendix \ref{IsingApp}. Therefore, we 
list below only   differences between our treatment of the two models
and their properties.

To start, we mention that Jordan-Wigner transformation  (\ref{SJW})
has to be   supplemented by 
$\sigma^y_i=\ii(c_i^\dag-c_i)\prod_{j<i}(1-2 c_j^\dag c_j)$. 
Then,  introducing
\be
\begin{aligned}
&A_k(g)=g-\tfrac{4}{3}\cos k+\tfrac{1}{3}\cos2k,\\
&B_k=\tfrac{2}{3}\sin k-\tfrac{1}{3}\sin2k,
\end{aligned}
\ee
we can concisely state that   (\ref{SHk}),
(\ref{thetaknew}), (\ref{thetak}), and (\ref{Sidt}), 
get now replaced by 
\be
\begin{aligned}
H = 2\sum_k \big[&( c_k^\dag  c_k - c_{-k}  c_{-k}^\dag)
A_k(g)\\
+&( c_k^\dag c_{-k}^\dag +  c_{-k} c_{k})B_k\big],
\end{aligned}
\label{SHEx}
\ee
\begin{align}
&\sin \theta_k = \frac{B_k}{
\sqrt{A_k^2(g)+B^2_k}
}, \\ 
&\cos\theta_k= \frac{A_k(g)}{
\sqrt{A_k^2(g)+B_k^2}
},
\end{align}
and
\be
\ii\frac{d}{dt}
\left(
\begin{array}{c}
v_k \\
u_k
\end{array}
\right)
=
2\left(
\begin{array}{cc}
A_k(g(t))    & -B_k     \\
-B_k & -A_k(g(t))
\end{array}
\right)\left(
\begin{array}{c}
v_k \\
u_k
\end{array}
\right),
\ee
respectively. 
The last difference is that 
the condition
for finite-size-independence of KZ dynamics now reads $\tau_Q\ll N^6$, because $z=3$
and $\nu=1/3$ in this model \cite{DebasisPRB2020}. 
We mention in passing that there are misprints 
in the expression for (\ref{SHEx})
in \cite{DebasisPRB2020}.

The rest of the discussion from  whole  Appendix \ref{IsingApp} 
identically characterizes  our studies of the extended $XY$ model.

\section{Scaling ansatz for  susceptibility of Ising model}
\label{ScalingApp}
To support the scaling ansatz for the susceptibility $\chi$ of the
Ising model, we start from consideration of 
\be
\tilde\chi=\frac{d\chi}{dg}.
\label{Stildechi}
\ee
Its  equilibrium singular part is given by 
\be
\tilde\chi^\eq_\sing\approx\frac{1}{\pi(1-g)},
\label{Sqwerty}
\ee
and so $\tilde\chi^\eq$ is algebraically divergent at the QCP. 
Applying to (\ref{Stildechi}) scaling ansatz (\ref{chit}),  we get
\be
\tilde\chi\approx\sqrt{\tau_Q}\tilde f((g-1)\sqrt{\tau_Q}),
\label{Schit}
\ee
which, when combined with (\ref{Stildechi}), leads to 
\be
\frac{d\chi}{dx}\approx \tilde f(x), \ \ x=(g-1)\sqrt{\tau_Q}.
\label{Szxcv}
\ee

Integrating (\ref{Szxcv}) over $x$, we get 
\be
\chi\approx h(x) + C,
\ee
where $h$ is a new scaling function.
To fix the $x$-independent $C$ term, we require that $\chi(x\ll-1)$ for the up
quench and $\chi(x\gg1)$ for the down quench are well-approximated by  
\be
\chi^\eq_\sing\approx -\frac{1}{\pi}\ln|g-1|.
\ee
Then, we note that by definition  scaling functions 
can depend on $\tau_Q$ only through their  argument. This  leads to
the conclusion that 
\be
\chi\approx h((g-1)\sqrt{\tau_Q})+\frac{\ln\tau_Q}{2\pi},
\label{SSS}
\ee
where $h(x\ll-1)$ for the up quench and $h(x\gg1)$ for the down quench are
well-approximated by $-\frac{1}{\pi}\ln|x|$.
After identification of
$h((g-1)\sqrt{\tau_Q})$ with $f((g(t)-g_c)/\hat g)$, (\ref{SSS})
matches (\ref{lnchit}).
We mention in passing that a factor of $2$, in the denominator of the second
term in (\ref{SSS}), can be traced back to $1+z\nu=2$.


\begin{thebibliography}{70}
\expandafter\ifx\csname natexlab\endcsname\relax\def\natexlab#1{#1}\fi
\expandafter\ifx\csname bibnamefont\endcsname\relax
  \def\bibnamefont#1{#1}\fi
\expandafter\ifx\csname bibfnamefont\endcsname\relax
  \def\bibfnamefont#1{#1}\fi
\expandafter\ifx\csname citenamefont\endcsname\relax
  \def\citenamefont#1{#1}\fi
\expandafter\ifx\csname url\endcsname\relax
  \def\url#1{\texttt{#1}}\fi
\expandafter\ifx\csname urlprefix\endcsname\relax\def\urlprefix{URL }\fi
\providecommand{\bibinfo}[2]{#2}
\providecommand{\eprint}[2][]{\url{#2}}

\bibitem[{Kib()}]{KibbleRev}
\bibinfo{note}{T. W. B. Kibble, Phys. Rep. {\bf 67},183 (1980).}

\bibitem[{Zur()}]{ZurekRev}
\bibinfo{note}{W. H. Zurek, Phys. Rep. {\bf 276}, 177 (1996).}

\bibitem[{\citenamefont{{Kibble}}(2007)}]{KibbleToday}
\bibinfo{author}{\bibfnamefont{T.}~\bibnamefont{{Kibble}}},
  \bibinfo{journal}{Phys. Today} \textbf{\bibinfo{volume}{60}},
  \bibinfo{pages}{47} (\bibinfo{year}{2007}).

\bibitem[{Jac({\natexlab{a}})}]{JacekAdv2010}
\bibinfo{note}{J. Dziarmaga, Adv. Phys. {\bf 59}, 1063 (2010).}

\bibitem[{Pol({\natexlab{a}})}]{PolkovnikovRMP2011}
\bibinfo{note}{A. Polkovnikov, K. Sengupta, A. Silva, and M. Vengalattore, Rev.
  Mod. Phys. {\bf 83}, 863 (2011).}

\bibitem[{del({\natexlab{a}})}]{delCampoReva}
\bibinfo{note}{A. del Campo, T. W. B. Kibble, and W. H. Zurek, J. Phys.:
  Condens. Matter {\bf 25}, 404210 (2013).}

\bibitem[{del({\natexlab{b}})}]{delCampoRevb}
\bibinfo{note}{A. del Campo and W. H. Zurek, Int. J. Mod. Phys. A {\bf 29},
  1430018 (2014).}

\bibitem[{Car()}]{Cardy1}
\bibinfo{note}{J. Cardy, {\it Scaling and Renormalization in Statistical
  Physics} (Cambridge University Press, Cambridge, 2002).}

\bibitem[{\citenamefont{{Coleman} and {Schofield}}(2005)}]{PiersNature2005}
\bibinfo{author}{\bibfnamefont{P.}~\bibnamefont{{Coleman}}} \bibnamefont{and}
  \bibinfo{author}{\bibfnamefont{A.~J.} \bibnamefont{{Schofield}}},
  \bibinfo{journal}{Nature} \textbf{\bibinfo{volume}{433}},
  \bibinfo{pages}{226} (\bibinfo{year}{2005}).

\bibitem[{Sac({\natexlab{a}})}]{Sachdev}
\bibinfo{note}{S. Sachdev, {\it {Q}uantum {P}hase {T}ransitions} (Cambridge
  University Press, 2011).}

\bibitem[{Con()}]{ContinentinoBook}
\bibinfo{note}{M. Continentino, {\it Quantum Scaling in Many-Body Systems: An
  Approach to Quantum Phase Transitions} (Cambridge University Press, 2nd
  edition, 2017).}

\bibitem[{Sac({\natexlab{b}})}]{SachdevToday}
\bibinfo{note}{S. Sachdev and B. Keimer, Phys. Today {\bf 64}, 29 (2011).}

\bibitem[{BDP({\natexlab{a}})}]{BDPRL2005}
\bibinfo{note}{B. Damski, Phys. Rev. Lett. {\bf 95}, 035701 (2005).}

\bibitem[{Dor({\natexlab{a}})}]{DornerPRL2005}
\bibinfo{note}{W. H. Zurek, U. Dorner, and P. Zoller, Phys. Rev. Lett. {\bf
  95}, 105701 (2005).}

\bibitem[{Jac({\natexlab{b}})}]{JacekPRL2005}
\bibinfo{note}{J. Dziarmaga, Phys. Rev. Lett. {\bf 95}, 245701 (2005).}

\bibitem[{\citenamefont{Yin et~al.}()\citenamefont{Yin, Qin, Lee, and
  Zhong}}]{ZhongArxiv2013}
\bibinfo{author}{\bibfnamefont{S.}~\bibnamefont{Yin}},
  \bibinfo{author}{\bibfnamefont{X.}~\bibnamefont{Qin}},
  \bibinfo{author}{\bibfnamefont{C.}~\bibnamefont{Lee}}, \bibnamefont{and}
  \bibinfo{author}{\bibfnamefont{F.}~\bibnamefont{Zhong}},
  \eprint{arXiv:1207.1602 (2013)}.

\bibitem[{\citenamefont{Hu et~al.}(2015)\citenamefont{Hu, Yin, and
  Zhong}}]{FanPRB2015}
\bibinfo{author}{\bibfnamefont{Q.}~\bibnamefont{Hu}},
  \bibinfo{author}{\bibfnamefont{S.}~\bibnamefont{Yin}}, \bibnamefont{and}
  \bibinfo{author}{\bibfnamefont{F.}~\bibnamefont{Zhong}},
  \bibinfo{journal}{Phys. Rev. B} \textbf{\bibinfo{volume}{91}},
  \bibinfo{pages}{184109} (\bibinfo{year}{2015}).

\bibitem[{\citenamefont{Huang and Yin}(2019)}]{HuangPRB2019}
\bibinfo{author}{\bibfnamefont{R.-Z.} \bibnamefont{Huang}} \bibnamefont{and}
  \bibinfo{author}{\bibfnamefont{S.}~\bibnamefont{Yin}},
  \bibinfo{journal}{Phys. Rev. B} \textbf{\bibinfo{volume}{99}},
  \bibinfo{pages}{184104} (\bibinfo{year}{2019}).

\bibitem[{Arn({\natexlab{a}})}]{ArnabSciRep2015}
\bibinfo{note}{S. Bhattacharyya, S. Dasgupta, and A. Das, Sci. Rep. {\bf 5},
  16490 (2015).}

\bibitem[{\citenamefont{Roy et~al.}(2017)\citenamefont{Roy, Moessner, and
  Das}}]{ArnabPRB2017}
\bibinfo{author}{\bibfnamefont{S.}~\bibnamefont{Roy}},
  \bibinfo{author}{\bibfnamefont{R.}~\bibnamefont{Moessner}}, \bibnamefont{and}
  \bibinfo{author}{\bibfnamefont{A.}~\bibnamefont{Das}},
  \bibinfo{journal}{Phys. Rev. B} \textbf{\bibinfo{volume}{95}},
  \bibinfo{pages}{041105(R)} (\bibinfo{year}{2017}).

\bibitem[{BDP({\natexlab{b}})}]{BDPRA2006}
\bibinfo{note}{B. Damski and W. H. Zurek, Phys. Rev. A {\bf 73}, 063405
  (2006).}

\bibitem[{\citenamefont{Tomka et~al.}(2018)\citenamefont{Tomka, Campos~Venuti,
  and Zanardi}}]{ZanardiPRA2018}
\bibinfo{author}{\bibfnamefont{M.}~\bibnamefont{Tomka}},
  \bibinfo{author}{\bibfnamefont{L.}~\bibnamefont{Campos~Venuti}},
  \bibnamefont{and} \bibinfo{author}{\bibfnamefont{P.}~\bibnamefont{Zanardi}},
  \bibinfo{journal}{Phys. Rev. A} \textbf{\bibinfo{volume}{97}},
  \bibinfo{pages}{032121} (\bibinfo{year}{2018}).

\bibitem[{Kol()}]{KolodrubetzPRL2012}
\bibinfo{note}{M. Kolodrubetz, B. K. Clark, and D. A. Huse, Phys. Rev. Lett.
  {\bf 109}, 015701 (2012).}

\bibitem[{Son()}]{SondhiPRB2012}
\bibinfo{note}{A. Chandran, A. Erez, S. S. Gubser, and S. L. Sondhi, Phys. Rev.
  B {\bf 86}, 064304 (2012).}

\bibitem[{Fra()}]{FrancuzPRB2016}
\bibinfo{note}{A. Francuz, J. Dziarmaga, B. Gardas, and W. H. Zurek, Phys. Rev.
  B {\bf 93}, 075134 (2016).}

\bibitem[{\citenamefont{Sadhukhan et~al.}(2020)\citenamefont{Sadhukhan, Sinha,
  Francuz, Stefaniak, Rams, Dziarmaga, and Zurek}}]{DebasisPRB2020}
\bibinfo{author}{\bibfnamefont{D.}~\bibnamefont{Sadhukhan}},
  \bibinfo{author}{\bibfnamefont{A.}~\bibnamefont{Sinha}},
  \bibinfo{author}{\bibfnamefont{A.}~\bibnamefont{Francuz}},
  \bibinfo{author}{\bibfnamefont{J.}~\bibnamefont{Stefaniak}},
  \bibinfo{author}{\bibfnamefont{M.~M.} \bibnamefont{Rams}},
  \bibinfo{author}{\bibfnamefont{J.}~\bibnamefont{Dziarmaga}},
  \bibnamefont{and} \bibinfo{author}{\bibfnamefont{W.~H.} \bibnamefont{Zurek}},
  \bibinfo{journal}{Phys. Rev. B} \textbf{\bibinfo{volume}{101}},
  \bibinfo{pages}{144429} (\bibinfo{year}{2020}).

\bibitem[{\citenamefont{Rossini and Vicari}(2020)}]{VicariPRR2020}
\bibinfo{author}{\bibfnamefont{D.}~\bibnamefont{Rossini}} \bibnamefont{and}
  \bibinfo{author}{\bibfnamefont{E.}~\bibnamefont{Vicari}},
  \bibinfo{journal}{Phys. Rev. Research} \textbf{\bibinfo{volume}{2}},
  \bibinfo{pages}{023211} (\bibinfo{year}{2020}).

\bibitem[{BDP({\natexlab{c}})}]{BDPRL2010}
\bibinfo{note}{B. Damski and W. H. Zurek, Phys. Rev. Lett. {\bf 104}, 160404
  (2010).}

\bibitem[{BDS()}]{BDSciRep2019}
\bibinfo{note}{O. A. Prośniak, M. Łącki, and B. Damski, Sci. Rep. {\bf 9},
  8687 (2019).}

\bibitem[{\citenamefont{Rams et~al.}(2018)\citenamefont{Rams, Sierant, Dutta,
  Horodecki, and Zakrzewski}}]{MarekPRX2018}
\bibinfo{author}{\bibfnamefont{M.~M.} \bibnamefont{Rams}},
  \bibinfo{author}{\bibfnamefont{P.}~\bibnamefont{Sierant}},
  \bibinfo{author}{\bibfnamefont{O.}~\bibnamefont{Dutta}},
  \bibinfo{author}{\bibfnamefont{P.}~\bibnamefont{Horodecki}},
  \bibnamefont{and}
  \bibinfo{author}{\bibfnamefont{J.}~\bibnamefont{Zakrzewski}},
  \bibinfo{journal}{Phys. Rev. X} \textbf{\bibinfo{volume}{8}},
  \bibinfo{pages}{021022} (\bibinfo{year}{2018}).

\bibitem[{\citenamefont{Suzuki}(1971)}]{SuzukiProgTeorPhys1971}
\bibinfo{author}{\bibfnamefont{M.}~\bibnamefont{Suzuki}},
  \bibinfo{journal}{Prog. Theor. Phys.} \textbf{\bibinfo{volume}{46}},
  \bibinfo{pages}{1337} (\bibinfo{year}{1971}).

\bibitem[{BDN()}]{BDNJP2008}
\bibinfo{note}{B. Damski and W. H. Zurek, New J. Phys. {\bf 10}, 045023
  (2008).}

\bibitem[{Mar()}]{MarekArxiv}
\bibinfo{note}{M. M. Rams, J. {Dziarmaga}, and W. H. {Zurek}, Phys. Rev. Lett.
  {\bf 123}, 130603 (2019).}

\bibitem[{Reg()}]{Regression}
\bibinfo{note}{One standard error, delivered by NonlinearModelFit function from
  \cite{Mathematica}, is provided in brackets in all our fitting results.}

\bibitem[{Lie()}]{Lieb1961}
\bibinfo{note}{E. Lieb, T. Schultz, and D. Mattis, Ann. Phys. (N.Y.) {\bf 16},
  407 (1961).}

\bibitem[{Pfe()}]{Pfeuty}
\bibinfo{note}{P. Pfeuty, Ann. Phys. {\bf 57}, 79 (1970).}

\bibitem[{BDJ()}]{BDJPA2014}
\bibinfo{note}{B. Damski and M. M. Rams, J. Phys. A {\bf 47}, 025303 (2014).}

\bibitem[{Pol({\natexlab{b}})}]{PolkovnikovPRB2005}
\bibinfo{note}{A. Polkovnikov, Phys. Rev. B {\bf 72}, 161201(R) (2005).}

\bibitem[{Ral()}]{RalfPRA2007}
\bibinfo{note}{S. Mostame, G. Schaller, and R. Sch\"utzhold, Phys. Rev. A {\bf
  76}, 030304(R) (2007).}

\bibitem[{Pol({\natexlab{c}})}]{PolkovnikovPRL2008}
\bibinfo{note}{R. Barankov and A. Polkovnikov, Phys. Rev. Lett. {\bf 101},
  076801 (2008).}

\bibitem[{Sen({\natexlab{a}})}]{SenPRB2009}
\bibinfo{note}{S. Mondal, K. Sengupta, and D. Sen, Phys. Rev. B {\bf 79},
  045128 (2009).}

\bibitem[{San({\natexlab{a}})}]{SantoroPRB2009}
\bibinfo{note}{D. Patan\`e, L. Amico, A. Silva, R. Fazio, and G. E. Santoro,
  Phys. Rev. B {\bf 80}, 024302 (2009).}

\bibitem[{Jac({\natexlab{c}})}]{JacekPRA2007}
\bibinfo{note}{L. Cincio, J. Dziarmaga, M. M. Rams, and W. H. Zurek, Phys. Rev.
  A {\bf 75}, 052321 (2007).}

\bibitem[{Sen({\natexlab{b}})}]{SenPRA2009}
\bibinfo{note}{K. Sengupta and D. Sen, Phys. Rev. A {\bf 80}, 032304 (2009).}

\bibitem[{Arn({\natexlab{b}})}]{ArnabPRB2010}
\bibinfo{note}{A. Das, Phys. Rev. B {\bf 82}, 172402 (2010).}

\bibitem[{Dut()}]{Dutta2015}
\bibinfo{note}{A. Dutta, G. Aeppli, B. K. Chakrabarti, U. Divakaran, T. F.
  Rosenbaum, and D. Sen, {\it Quantum Phase Transitions in Transverse Field
  Spin Models: From Statistical Physics to Quantum Information} (Cambridge
  University Press, 2015).}

\bibitem[{San({\natexlab{b}})}]{SantoroJstat2015}
\bibinfo{note}{A. Russomanno, S. Sharma, A. Dutta, and G. E. Santoro, J. Stat.
  Mech. (2015) P08030.}

\bibitem[{BDp()}]{BDproceedings}
\bibinfo{note}{B. Damski, Fidelity approach to quantum phase transitions in
  quantum Ising model, in {\it Quantum Criticality in Condensed Matter:
  Phenomena, Materials and Ideas in Theory and Experiment}, edited by J.
  Jedrzejewski (World Scientific, Singapore, 2015), pp. 159–182;
  arXiv:1509.03051.}

\bibitem[{Pus()}]{PuskarovSciP2016}
\bibinfo{note}{T. Puskarov and D. Schuricht, SciPost Phys. {\bf 1}, 003
  (2016).}

\bibitem[{Apo()}]{ApollaroSciRep2017}
\bibinfo{note}{S. Lorenzo, J. Marino, F. Plastina, G. M. Palma, and T. J. G.
  Apollaro, Sci. Rep. {\bf 7}, 5672 (2017).}

\bibitem[{Mic({\natexlab{a}})}]{MichalOne}
\bibinfo{note}{M. Bia{\l}o{\'{n}}czyk and B. Damski, J. Stat. Mech. (2018)
  073105.}

\bibitem[{Ado()}]{AdolfoPRL2018}
\bibinfo{note}{A. del Campo, Phys. Rev. Lett. {\bf 121}, 200601 (2018).}

\bibitem[{Mic({\natexlab{b}})}]{MichalTwo}
\bibinfo{note}{M. Bia{\l}o{\'{n}}czyk and B. Damski, J. Stat. Mech. (2020)
  013108.}

\bibitem[{Lac()}]{LackiJSTAT2017}
\bibinfo{note}{M. {\L}{\k{a}}cki and B. Damski, J. Stat. Mech. (2017) 103105.}

\bibitem[{Tur()}]{TurbanJPA2007}
\bibinfo{note}{T. Platini, D. Karevski, and L. Turban, J. Phys. A: Math. Theor.
  {\bf 40}, 1467 (2007).}

\bibitem[{Dor({\natexlab{b}})}]{DornerPRS2008}
\bibinfo{note}{W. H. Zurek and U. Dorner, Phil. Trans. R. Soc. A {\bf 366},
  2953 (2008).}

\bibitem[{\citenamefont{Verstraete et~al.}(2008)\citenamefont{Verstraete, Murg,
  and Cirac}}]{VerstraeteAdvPhys2008}
\bibinfo{author}{\bibfnamefont{F.}~\bibnamefont{Verstraete}},
  \bibinfo{author}{\bibfnamefont{V.}~\bibnamefont{Murg}}, \bibnamefont{and}
  \bibinfo{author}{\bibfnamefont{J.}~\bibnamefont{Cirac}},
  \bibinfo{journal}{Adv. Phys.} \textbf{\bibinfo{volume}{57}},
  \bibinfo{pages}{143} (\bibinfo{year}{2008}).

\bibitem[{\citenamefont{{Schollw{\"o}ck}}(2011)}]{SCHOLLWOCKAnnals2011}
\bibinfo{author}{\bibfnamefont{U.}~\bibnamefont{{Schollw{\"o}ck}}},
  \bibinfo{journal}{Ann. Phys.} \textbf{\bibinfo{volume}{326}},
  \bibinfo{pages}{96} (\bibinfo{year}{2011}).

\bibitem[{\citenamefont{{Or{\'u}s}}(2014)}]{OrusAnnals2014}
\bibinfo{author}{\bibfnamefont{R.}~\bibnamefont{{Or{\'u}s}}},
  \bibinfo{journal}{Ann. Phys.} \textbf{\bibinfo{volume}{349}},
  \bibinfo{pages}{117} (\bibinfo{year}{2014}).

\bibitem[{Por()}]{PorrasPRL2004}
\bibinfo{note}{D. Porras and J. I. Cirac, Phys. Rev. Lett. {\bf 92}, 207901
  (2004).}

\bibitem[{Kor()}]{KorenblitNJP2012}
\bibinfo{note}{S. Korenblit {\it et al.}, New J. Phys. {\bf 14}, 095024
  (2012).}

\bibitem[{Lew()}]{LewensteinAdv}
\bibinfo{note}{M. Lewenstein, A. Sanpera, V. Ahufinger, B. Damski, A. Sen De,
  and U. Sen, Adv. Phys. {\bf 56}, 243 (2007).}

\bibitem[{Blo()}]{BlochSci2017}
\bibinfo{note}{C. Gross and I. Bloch, Science {\bf 357}, 995 (2017).}

\bibitem[{Sch()}]{SchaussQST2018}
\bibinfo{note}{P. Schauss, Quantum Sci. Technol. {\bf 3}, 023001 (2018).}

\bibitem[{Luk({\natexlab{a}})}]{LukinNature2017}
\bibinfo{note}{H. Bernien {\it et al.}, Nature {\bf 551}, 579 (2017).}

\bibitem[{Mon()}]{MonroeNature2017}
\bibinfo{note}{J. Zhang, G. Pagano, P. W. Hess, A. Kyprianidis, P. Becker, H.
  Kaplan, A. V. Gorshkov, Z.-X. Gong, and C. Monroe, Nature {\bf 551}, 601
  (2017).}

\bibitem[{Luk({\natexlab{b}})}]{LukinNature2019}
\bibinfo{note}{A. Keesling {\it et al.}, Nature {\bf 568}, 207 (2019).}

\bibitem[{LMG()}]{LMGexp}
\bibinfo{note}{V. Makhalov, T. Satoor, A. Evrard, T. Chalopin, R. Lopes, and S.
  Nascimbene, Phys. Rev. Lett. {\bf 123}, 120601 (2019).}

\bibitem[{Rec()}]{Recipes}
\bibinfo{note}{W. H. Press, S. A. Teukolsky, W. T. Vetterling, and B. P.
  Flannery, {\it Numerical recipes in C. The art of scientific computing}
  (Cambridge University Press, 2nd edition, 1992).}

\bibitem[{Mat()}]{Mathematica}
\bibinfo{note}{Wolfram Research, Inc., Mathematica, Version 12.0, Champaign, IL
  (2019).}

\end{thebibliography}

\end{document}